\documentclass[10pt,aps,prd,twocolumn,showpacs,nofootinbib,preprintnumbers,a4paper]{revtex4-1}
\pdfoutput=1
\usepackage{fullpage}
\usepackage{amsmath}
\usepackage[pdftex]{graphicx}
\usepackage{tikz}
\usetikzlibrary{mindmap,trees}
\usetikzlibrary{decorations.shapes}
\usepackage[caption=false]{subfig}
\usepackage[english]{babel}
\usepackage{amssymb}
\usepackage{amsfonts}
\usepackage{bbold}
\usepackage{bm}
\usepackage{color}
\usepackage{xcolor}
\usepackage{dcolumn}
\usepackage[colorlinks]{hyperref}
\usepackage{cancel}
\usepackage{units}
\usepackage[normalem]{ulem}
\hypersetup{
pdfauthor={},
pdftitle={CKM and PMNS from discrete symmetry}
}

\newcommand{\cAfour}{\cite{Babu:2005uq,Ma:2004qy,Babu:2003fk,Ma:2001lr,He:2006dk,Altarelli:2006qy,Altarelli:2005uq}}

\newcommand{\GAP}{\texttt{GAP}}

\newcommand{\diag}{\ensuremath{\mathrm{diag}}}

\newcommand{\SU}[1]{\ensuremath{\mathrm{SU}(#1)}}

\newcommand{\U}[1]{\ensuremath{\mathrm{U}(#1)}}

\begin{document}
\allowdisplaybreaks[1]
\widetext

\title{Quark and Leptonic Mixing Patterns from the Breakdown of a Common Discrete Flavor Symmetry}
\vspace*{10mm}
\author{Martin Holthausen}
\email{martin.holthausen@mpi-hd.mpg.de}
\author{Kher Sham Lim}
\email{khersham.lim@mpi-hd.mpg.de}
\affiliation{Max-Planck-Institut f\"{u}r Kernphysik, Saupfercheckweg 1, 69117 Heidelberg, Germany}

\begin{abstract}
Assuming the Majorana nature of neutrinos, we recently performed a scan of leptonic mixing patterns derived from finite discrete groups of order less than 1536. Here we show that the 3 groups identified there as giving predictions close to experiment, also contain another class of abelian subgroups that predict an interesting leading order quark mixing pattern where only the Cabibbo angle is generated at leading order. We further broaden our study by assuming that neutrinos are Dirac particles and find 4 groups of order up to 200 that can predict acceptable quark and leptonic mixing angles. Since large flavor groups allow for a multitude of leading order mixing patterns, we define a measure that is suitable to compare the predictivity of a given flavor group taking this fact into account. We give the result of this measure for a wide range of discrete flavor groups and identify the group $(Z_{18}\times Z_6)\rtimes \mathcal{S}_3$  as being most predictive in the sense of this 
measure. We 
further discuss 
alternative measures and their implications.
\end{abstract}
\maketitle

\section{Introduction \label{sec:outline}}

With the discovery of a Higgs-like resonance at 126 GeV, the Standard Model appears to be complete and from a purely phenomenological standpoint no new physics seems to be required up to a very large scale, e.g. up to the Planck scale. While the description of gauge interactions in the Standard Model is quite economical (requiring only 3 parameters), the fact that there are three generations of fermions is not explained in the Standard Model and necessitates the introduction of many additional parameters into the model. Furthermore, these flavor parameters show certain structures that may suggest a deeper explanation: the quark sector exhibits a strongly hierarchical mass spectrum and small mixing angles while the lepton sector is less hierarchical and has larger mixing angles.

There have been many attempts in the literature to try and explain these structures using symmetries that act on the different families. Here we focus on models with non-abelian discrete flavor symmetries, which are known to be able to describe the large mixing angles of the lepton sector. The general setup of such models is as follows: a discrete flavor group is broken to different subgroups in the charged lepton and neutrino sectors and the mismatch between the two subgroups allows one to predict the PMNS matrix (up to permutations of rows and columns of the matrix) \cite{Lam:2007fk,Lam:2008sh,Lam:2008fj,Toorop:2011jn,de-Adelhart-Toorop:2012fv,Lam:2012he}. It should be noted that these predictions will in general be slightly perturbed by the inclusion of higher dimensional operators and renormalization group running of parameters, which we will subsume under next-to-leading order (NLO) effects. 
 
Recently, we performed a comprehensive scan of leptonic mixing parameters that can be obtained from remnant symmetries which form a group of size smaller than 1536 \cite{Holthausen:2012wt}. We identified the groups $\Delta(6\cdot 10^2)$, $(Z_{18}\times Z_6)\rtimes \mathcal{S}_3$ and $\Delta(6\cdot 16^2)$ as being the only ones that may reproduce the experimentally favored mixing angles. All three groups are either of the form $\Delta(6\cdot n^2)$ \cite{Escobar:2009mz}, or a subgroup of such a group (see \cite{King:2013vna} for a recent study of these symmetry groups). 

\begin{figure}[bt]
\centering
\begin{tikzpicture}[scale=2]
\draw (0,0) node (Gf) {{\Large $G_f$}};
\draw (-1,1) node (Ge) {{\Large $G_e$}};
\draw (1,1) node (Gnu) {{\Large $G_\nu$}};
\draw (-1,-1) node (Gu) {{\Large $G_u$}};
\draw (1,-1) node (Gd) {{\Large $G_d$}};
\draw[->] (Gf) to (Ge);
\draw[->] (Gf) to (Gnu);
\draw[->] (Gf) to (Gu);
\draw[->] (Gf) to (Gd);
\draw[<->, in=170,out=10,color=blue!50] (Ge) to node[above, color=blue!50,yshift=1ex]{PMNS}  (Gnu);
\draw[<->, in=190,out=350,, color=blue!50] (Gu) to node[above, color=blue!50,yshift=1ex]{CKM}  (Gd);
\end{tikzpicture}
\caption{Sketch of the setup considered in this paper. Different subgroups of the flavor group $G_f$ emerge as remnant symmetries of the mixing matrices. The mismatch of these groups creates quark and lepton mixing.}
\label{fig:sketch}
\end{figure}
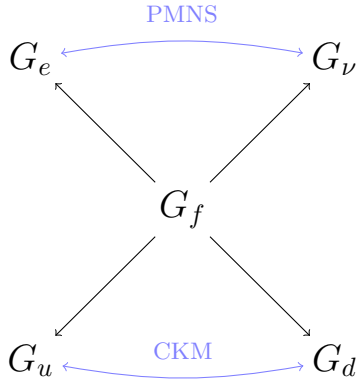

In this work, we study the question if also the quark mixing angles may be obtained to leading order (LO) as a result of mismatched remnant symmetries of non-abelian discrete groups. Since the Cabibbo angle $\theta_c$ is roughly of similar size as the reactor mixing angle 
$$
\theta_{13}\simeq \frac{\theta_{c}}{\sqrt{2}}\simeq 9.2^\circ
$$
it would be interesting to obtain patterns in which all leptonic plus the Cabibbo angle are produced at leading order as a result of remnant symmetries. Since the other angles are smaller, it is prudent to assume them to be a result of NLO corrections. It turns out that if one assigns the left-handed quarks to the same 3-dimensional representations (of the same groups) that were found to be interesting for leptonic mixing, such an interesting quark mixing pattern may be derived. Especially the group $(Z_{18}\times Z_6)\rtimes \mathcal{S}_3$ seems particularly promising, giving a Cabibbo angle of $\sin \theta_c=0.259$. In this setup the origin of the different patterns for the leptonic and quark sectors thus stems from the different remnant symmetries to which the original group is broken in the respective sectors, as is depicted in Fig.~\ref{fig:sketch}. 
The subgroups that give rise to an acceptable LO Cabibbo angle can be systematically parametrized and we discuss some group theoretical aspects of the remnant group structure.
We then broaden our discussion by giving up on the assumption that neutrinos should be Majorana particles, for which case we perform a scan of finite discrete groups up to the order of 200 with the help of the computer algebra program \GAP~\cite{GAP4:2011,REPSN:2011,SmallGroups:2011,SONATA:2003}. 

In the final chapter of the paper we discuss the usefulness of large flavor groups more generally. It should be clear that if one considers a very large flavor group virtually any mixing pattern may be realized. If one would break the group $\SU{3}$ down to discrete remnant groups, for example, the requirement of a breakdown to subgroups loses all predictivity. The question one might now ask oneself is the following: which setup is more predictive, the case of a small group (such as $A_4$ \cAfour~ or $S_4$ \cite{Lam:2007fk,Lam:2008fj,Lam:2008sh}) with large NLO corrections or a larger group with smaller NLO corrections. Under the assumption that NLO corrections are randomly drawn (which seems fine for many models) statistical arguments (\`{a}	 la anarchy) lead us to propose a measure of the predictive power of a chosen group. 

\section{PMNS and CKM matrices from remnant symmetries \label{sec:mixing}}

Here we briefly review the setup we are using to obtain the mixing matrices from remnant symmetries.

Lepton mixing can be obtained from a flavor symmetry group via its breaking to remnant symmetries in the charged lepton and neutrino masses respectively. The CKM matrix can be derived in an analogous way using this method. The only difference is that usually different remnant symmetries are left of the up- and down-type quarks mass matrices. This is usually achieved in concrete models via spontaneous symmetry breaking of flavon fields in some vacuum alignment configurations. As in Ref.~\cite{Holthausen:2012wt} we do not consider the breaking mechanisms or models to achieve such vacuum configuration, but rather we want to find discrete symmetry groups that contain the residual symmetry groups that can give rise to LO prediction of PMNS and CKM matrices.

In this section we first assume that neutrinos are Majorana particles. The PMNS and CKM matrices are defined as
\begin{align}
U_{\mathrm{PMNS}}=V_e^\dagger V_\nu \, , \quad U_{\mathrm{CKM}}=V_d^\dagger V_u
\end{align}
where the unitary matrices $V_s$ and $V_\nu$ diagonalize the mass matrices
\begin{align}
V_s^T M_s M_s^\dagger V_s^*=\diag(m_{\mathrm{I}}^2,m_{\mathrm{II}}^2,m_{\mathrm{III}}^2)
\end{align}
and
\begin{align}
V_\nu^T M_\nu V_\nu=\diag(m_1, m_2, m_3). 
\end{align}
We denote the symbol $s \in \{e,d,u\}$ and the numeral $\mathrm{I}\in \{e,d,u\}$, $\mathrm{II}\in \{ \mu,s,c \}$ and $\mathrm{III} \in \{\tau,b,u \}$. The mass matrices are defined as $\mathcal{L}=e^T M_e e^c+ \frac{1}{2}\nu^T M_\nu \nu+d^T M_d d^c + u^T M_u u^c$. We assume that there is a discrete symmetry group $G_f$ under which the left-handed lepton doublets $L=(\nu,e)$ transform under a faithful unitary 3-dimensional representation $\rho:G_f\rightarrow GL(3,\mathbb{C})$:
\begin{align}
L \rightarrow \rho(g) L,\quad g \in G_f.  
\end{align}
Analogue we assume that there is a discrete symmetry group $G_Q$ under which the left-handed quark doublets $Q=(u,d)$ transforms:
\begin{align}
Q \rightarrow \rho(g) Q,\quad g \in G_Q.
\label{eq:qtrafo}
\end{align}
Since all the quark and lepton masses are different, these flavor symmetries has to be broken into two set of different subgroups, i.e. $\{G_e,G_\nu\}$ for the leptonic sector and $\{G_d, G_u\}$ for the quark sector. In general the generators of $G_d$ and $G_u$ only generate the group $G_Q$ which is a proper subgroup of $G_f$, hence we only consider a direct breaking of $G_f$ into residual symmetries $G_d$ and $G_u$ as we would like to find a common discrete group $G_f$ that can simultaneously predict the LO PMNS and CKM matrix\footnote{This possibility is also briefly discussed in Ref.~\cite{de-Adelhart-Toorop:2012fv}.}. Within a set of the residual subgroups $\{ G_e, G_\nu\}$, the intersection between the subgroups in the set is trivial as we would like to predict 3 different mixing angles in the leptonic sector. This condition is however relaxed for $\{ G_d,G_u \}$ as we do not find any groups that predict 3 different quark mixing angles at LO. Subgroups from different set, e.g. $G_e$ and $G_d$ can have 
non-trivial intersection. The mass matrix for each sector exhibits a residual symmetry, satisfying
\begin{align}
\rho(g_s)^T M_s M_s^\dagger \rho(g_s)^*=M_s M_s^\dagger,\quad g_s \in G_s  
\end{align}
and
\begin{align}
\rho(g_\nu)^T M_\nu \rho(g_\nu)=M_\nu, \quad g_\nu \in G_\nu. 
\end{align}
The residual subgroups $\{G_e,G_\nu \}$ and $\{ G_d,G_u\}$ must be abelian due to the experimental fact that all the masses of quarks and leptons are distinguishable. The 3-dimensional irreducible representation of the residual subgroups cannot be decomposed into three inequivalent 1-dimensional representations had they possess a non-abelian character\footnote{See Ref.~\cite{Hernandez:2013vya} for the case where neutrinos masses are degenerate.}. For Majorana neutrinos the residual subgroup is given by the Klein group $Z_2\times Z_2$, while $G_s$ can be any abelian subgroups of $G_f$ with order $n\ge3$. Once the generators of all the subgroups are specified in a certain representation, the mixing pattern of quark sector and leptonic sector can be determined via the unitary matrices $\Omega_s$ and $\Omega_\nu$ satisfying
\begin{align}
\Omega_{s,\nu}^\dagger \rho(g_{s,\nu}) \Omega_{s,\nu}= \rho(g_{s,\nu})_{\diag}.
\end{align}
The unitary matrix $\Omega_s$ and $\Omega_\nu$ are determined up to permutations of columns and also a diagonal phase matrix. The PMNS and CKM matrix are then determined by
\begin{align}
U_{\mathrm{PMNS}}=\Omega_e^\dagger \Omega_\nu \, , \quad U_{\mathrm{CKM}}=\Omega_d^\dagger \Omega_u.
\end{align}
which are unique up to the permutations of rows and columns. The Dirac CP phases of the PMNS and CKM matrices can also be determined from this method.

\section{Mixing pattern from common discrete symmetries\label{sec:result}}
As shown in Ref.~\cite{Holthausen:2012wt}, a scan of finite discrete groups with order less than 1536 yields only 3 interesting groups that give LO leptonic mixing patterns which lie within 3-sigma of current best fit. These 3 groups, namely $\Delta(6\cdot 10^2)$, $(Z_{18}\times Z_6)\rtimes \mathcal{S}_3$ and $\Delta(6\cdot 16^2)$, provide a good starting point to search for residual groups that can yield an acceptable CKM matrix at LO. By searching the abelian subgroups contained in these 3 groups, we obtain the CKM matrix at LO in the following form:
\begin{align}
U_{\mathrm{CKM}}=\left(\begin{array}{ccc}
\cos \tilde{\theta}&\sin \tilde{\theta}&0\\
-\sin \tilde{\theta}&\cos \tilde{\theta}&0\\
0&0&1
\end{array}\right).
\label{eq:ckmform}
\end{align}
 The values of $\sin \tilde{\theta}$ are given in Table \ref{tab:mixinga} and the form may be compared to best fit values of the CKM matrix \cite{Beringer:1900zz}
\begin{align}
U_{\mathrm{CKM}}\simeq\left(\begin{array}{ccc}
0.974&0.225&0.004\\
0.225&0.973&0.041\\
0.009&0.040&0.999
\end{array}\right),
\label{eq:ckm-values}
\end{align}
indicating that NLO corrections of the order of $U_{cb}\sim \lambda_c^2 \sim 0.04$ are needed, which is to be contrasted with the case of $A_4$, for example, where $U_{\mathrm{CKM}}=\mathbb{1}_3$ at LO and NLO corrections therefore have to be of the size $U_{cs}\sim \lambda_c\equiv \sin \theta_c \sim 0.22$. Since there is no mixing between all three generations in Eq. (\ref{eq:ckmform}) the CKM CP phase in undetermined in this setup and will be a result of NLO corrections.

Before we discuss the results of Table \ref{tab:mixinga}, it is useful to recall \cite{Holthausen:2012wt}  that the groups in Table \ref{tab:mixinga} may be defined as being generated by the generators $S$, $T$ and $U(n,k)$, using the faithful irrep $\rho: \{ S,T, U(n,k)\}\rightarrow \{S_3, T_3, U_3(n,k)\}$ with  
\begin{align}
T_3\equiv \left(\begin{array}{ccc}
0&1&0\\
0&0&1\\
1&0&0
\end{array}\right),\quad
S_3\equiv \left(\begin{array}{ccc}
1&0&0\\
0&-1&0\\
0&0&-1
\end{array}\right)
\label{eq:S3-def}
\end{align}
and 
\begin{align}
 U_3(n,k) \equiv-\left( \begin{array}{ccc} 1&0&0\\0&0&z_{n,k}\\0&z_{n,k}^*&0 \end{array}\right)
 \label{eq:Un-def}
\end{align}
with $z_{n,k}=e^{2\pi i k/n}$, $n,k\in \mathbb{N}$. In the leptonic sector if one uses $G_e=\langle T\rangle\cong Z_3$ and $G_\nu=\langle S, U(n,k)\rangle\cong Z_2\times Z_2$ one gets the TM2-like mixing matrix \cite{Holthausen:2012wt} 
\begin{align}
U_{\mathrm{PMNS}}=U_{\mathrm{HPS}}U_{13}(\theta=\frac{1}{2} \arg(z))
\label{eq:final} 
\end{align}
with the 1-3 rotation matrix defined as
\begin{align}
 U_{13}(\theta)=\left(
\begin{array}{ccc}
 \cos \theta & 0 & \sin \theta \\
0&1&0\\
-\sin\theta &0&\cos \theta
\end{array}
\right).
\label{eq:U13def}
\end{align}

In the quark sector we found two different types of solutions corresponding to different conserved subgroups. From the form \eqref{eq:ckmform} of the LO CKM matrix it is already clear that the intersection between $G_u$ and $G_d$ has to be non-vanishing, otherwise there would be full 3 by 3 mixing (as in the leptonic case). The generator of the intersection can in principle be any generator, but we will always take $S$ for concreteness. As a result of the scan, we found 2 types of mixing patterns
\begin{itemize}
\item type A: \begin{align*}G_d&=\langle S,U(n,p)\rangle\cong Z_2\times Z_2,\\ G_u&=\langle (ST)^2TU(n,m) \rangle \cong Z_4\end{align*}
\item type B: \begin{align*}G_d&=\langle S,U(n,p)\rangle\cong Z_2\times Z_2,\\ G_u&=\langle S,(U(n,m)T^2)^2 (U(n,m)T)^2 U(n,m)\rangle \\ &\cong Z_2\times Z_2\end{align*}
\end{itemize}
Both left-handed quarks and leptons may be assigned to the same representation, which provides a possibility for model building of flavor symmetry in the context of Grand Unified Theories.

\begin{table}
 \centering
 \begin{tabular}{cclccccc}
\hline
$n$ & &$G_f$&&\GAP-Id && $\sin \tilde{\theta}$  &type \\
\hline 5 && $\Delta(6\cdot 10^2)$ && $[600,179]$ &&  $0.156$&A \\
&&&&&&  $0.309$ &B\\
9 && $(Z_{18}\times Z_6)\rtimes \mathcal{S}_3$ && $[648,259]$ &&  $0.259$&A \\
16&& $\Delta(6\cdot 16^2)$ && n.a. &&  $0.195$ &A \\ \hline
\end{tabular}
\caption{LO Cabibbo angles $\sin \tilde{\theta}$  which are compatible with experimental results generated by flavor groups up to order 1536. Type A and B refers to different residual symmetries (see text).}
\label{tab:mixinga}
\end{table}

Let us first discuss the case of type A. The LO CKM matrix of Eq.~\eqref{eq:ckmform} results from the breakdown of $G_f$ down to $G_d=\langle S,U(n,m)\rangle\cong Z_2\times Z_2 $ and $G_u=\langle (ST)^2TU(n,p) \rangle \cong Z_4$. Note that $((ST)^2TU(n,p))^2=S$ is an element of both $G_d$ and $G_u$.

The generator of the group $G_u$ is given by
\begin{align}
R_3(n,p)&\equiv\rho((ST)^2TU(n,p)) \nonumber \\
&= \left( \begin{array}{ccc} 1&0&0\\0&0&-z_{n,p}\\0&z_{n,p}^*&0 \end{array}\right)
\label{eq:Rn-def}
\end{align}
with $z$ defined in Eq.~\eqref{eq:Un-def}. Note that typically one needs to choose a different $n$-th root in Eq.~\eqref{eq:Un-def} and Eq.~\eqref{eq:Rn-def} in order to obtain experimentally acceptable PMNS and CKM matrices. For example, if we choose the $m$-th of $n$-th root $z$ in Eq.~\eqref{eq:Un-def} and $p$-th of $n$-th root $z$ in Eq.~\eqref{eq:Rn-def}, the product of the unitary matrix 
\begin{align}
\Omega_u = \frac{1}{\sqrt{2}}\left( 
\begin{array}{ccc} 
0&0&\sqrt{2} \\
i e^{2\pi i p/n} & -ie^{2\pi i p/n} & 0\\
1&1&0 
\end{array}\right)
\label{eq:omegau}
\end{align}
that diagonalizes $R_3(n,p)$ with the unitary matrix
\begin{align}
\Omega_d = \frac{1}{\sqrt{2}}\left( 
\begin{array}{ccc} 
0&0&\sqrt{2} \\
e^{2\pi i m/n} & -e^{2\pi i m/n} & 0\\
1&1&0 
\end{array}\right)
\label{eq:omegad}
\end{align}
that diagonalizes $S_3$ and $U_3(n,m)$ simultaneously will generate LO CKM matrix
\begin{align}
&U_{\mathrm{CKM}}=\Omega_d^\dagger \Omega_u  \\ 
&= \frac{1}{2}\left( 
\begin{array}{ccc} 
1+ie^{-2\pi i (m-p)/n} & 1-ie^{-2\pi i (m-p)/n} & 0\\
1-ie^{-2\pi i (m-p)/n} & 1+ie^{-2\pi i (m-p)/n} & 0\\
0&0&2
\end{array}\right)\nonumber
\label{eq:omegax}
\end{align}
or 
\begin{align}
\sin \tilde{\theta} =\frac{1}{2} \sqrt{2-2 \sin \left(\frac{2 \pi  (m-p)}{n}\right)}
\end{align}
The interesting cases quotes in Table \ref{tab:mixinga} correspond to $(n=5, p=1, m=2)$, $(n=9, p=1, m=4)$ and $(n=16, p=1, m=2)$, respectively. Since $G_u$ and $G_d$ have a non-trivial intersection, the group generated by the elements of $G_u$ and $G_d$ is not the full flavor group $G_f$. Rather it is a subgroup of $\U{2}$, depending on the values of $n$, $p$ and $m$. The groups generated by these remnant symmetries are isomorphic to $(Z_{10}\times Z_2)\rtimes Z_2$, $(Z_6\times Z_2)\rtimes Z_2$ and $QD_{32}$ (the quasidihedral group of order 32), respectively\footnote{See Ref.~\cite{Ishimori:2010zr} for a review on the type of groups above.}.

The case of type B is analogous and one finds
\begin{align}
\sin \tilde{\theta} =\left\vert \cos \left(\frac{\pi  (m-4 p)}{n}\right)\right\vert,
\end{align}
where the case quoted in Table \ref{tab:mixinga} corresponds to $(n=5, p=1, m=1)$, which generates $D_{20}$, the dihedral group of size $20$. Dihedral groups of this type have been considered before as an explanation of the LO Cabbibo angle \cite{Blum:2009nh} as we will comment on in more detail below.

\begin{table*}[tb]
 \centering
 \begin{tabular*}{1\textwidth}{@{\extracolsep{\fill} }llll|llll}
 \hline
 $G_f$ & \GAP-Id &$\{ G_e,G_\nu \}$ & $\{ G_d,G_u \}$ & $\sin^2(\theta_{12})$ & $\sin^2(\theta_{13})$ & $\sin^2(\theta_{23})$ & $\sin \tilde{\theta}$  \\
 \hline
 $\Delta(6\cdot 5^2)$ & $[150,5]$ & $\{ Z_{10},Z_{3}\}$ & $\{Z_{10},Z_{10} \}$ & 0.3428 & 0.0289 & 0.6217 & 0.309 \\
 &&&& 0.3428 & 0.0289 & 0.3794 &  \\
 \hline
 $\Sigma(3\cdot 3^3)\rtimes Z_2$ & $[162,10]$ & $\{ Z_6,Z_9 \}$ & $\{ Z_6,Z_6 \}$ & 0.3403 & 0.0202 & 0.6013 & 0.5 \\
 $(Z_9\times Z_3)\rtimes \mathcal{S}_3$ & $[162,12]$ & $\{ Z_{18},Z_9 \}$ & $\{ Z_{18},Z_{18} \}$ & 0.3403 & 0.0202 & 0.3996 &  \\
 & $[162,14]$ & $\{ Z_{18},Z_3 \}$ & $\{ Z_{18},Z_{18} \}$ &&&& \\ \hline
 \end{tabular*}
\caption{Lepton mixing parameters and LO CKM entries predicted by finite discrete groups with order 150 and 162. The smallest generators for $G_e,G_\nu$ and $G_d,G_u$ that predict the quark and leptonic mixing angles on the right columns are listed.}
\label{tab:dirac}
\end{table*}

To recapitulate: we have seen that the structure of the LO CKM mixing \eqref{eq:ckmform} may be understood as a result of symmetry breaking down to the subgroups of type A and type B. The groups $\Delta(6\cdot 10^2)$, $(Z_{18}\times Z_6)\rtimes \mathcal{S}_3$ and $\Delta(6\cdot 16^2)$ are of the form $(Z_n \times Z_{n'}) \rtimes \mathcal{S}_3$, where $Z_n\cong \langle (ST)^2 (U(n,1)T)^4 T \rangle $, $Z_{n'}\cong \langle STSU(n,1)T^2U(n,1)T^2U(n,1)TU(n,1) \rangle $ and $\mathcal{S}_3=\langle R' ,T^2 R' T R'\rangle $, where $R'$ is short for $R'=(U(n,1)T^2)^2 (U(n,1)T)^2 U(n,1)$, one of the generators of $G_u$ in type B. Using this structure, the interested reader may figure out the origin of the remnant symmetries for the general case. However, from the 3 concrete cases we have studied in detail, we can infer the origin of subgroups of type A and B from a group theoretical perspective. The subgroups of type A and B consist of groups of type $(Z_m \times Z_{m'}) \rtimes Z_2$, which are always subgroups of $(Z_n \times Z_{n'}) \rtimes \mathcal{S}_3$ with 
$n^{(')}\ge m^{(')}$ (One of the $Z_m$ can be trivial). Therefore the 1-2 mixing structure of Eq.\eqref{eq:ckmform} is a by-product that we obtain for free from the leptonic flavor symmetry. It is also interesting to imagine the possibility that $G_Q$ is not a subgroup of $G_L=\langle G_e, G_\nu\rangle $ but that they rather be subgroups of yet larger group $G_f=\langle G_Q,G_L\rangle$. However from all the discrete groups that predict the experimentally favored values, $G_Q$ is always a subgroup of $G_L\equiv G_f$, hence an extension to larger group will not yield new interesting predictions.

In this study we only considered groups that are interesting because they give a good LO description of leptonic mixing. If one does not require the quark flavor group to be identical to the lepton flavor group, one may search for a flavor group $G_Q$ that predicts an adequate CKM matrix, independent of the leptonic flavor group $G_f$ \cite{Blum:2009nh,Lam:2007qc,Blum:2007nt,Hagedorn:2012pg}. 

As we have noted above, the group generated by $G_u$ and $G_d$ is not the full flavor group $G_f$ but a smaller group $G_Q$. The 3-dimensional representation $\textbf{3}$ of $G_f$ is decomposed into $\textbf{3}=\textbf{2}+\textbf{1}$, where the 2-dimensional representation $\textbf{2}$ of $G_Q$ generates the Cabibbo angle. This is very similar (at least for the symmetry breaking of type B) to models where one assigns the first two quark generations to a 2-dimensional representation of a dihedral group $D_n$\footnote{The symmetry breaking of type A might be viewed as a generalisation thereof if one replaces dihedral with the involved groups, e.g. quasidihedral.}. One may therefore view the groups discussed here as completions of groups that only discuss the quark sector.

Let us recapitulate on the search for unified discrete symmetry from the group theoretical approach. We started our scan for groups that can yield sizeable leptonic mixing patterns with the assumption of Majorana neutrinos. From over a million groups, only groups of type $(Z_n \times Z_{n'}) \rtimes \mathcal{S}_3$ are found to be interesting and such groups contain subgroups, which allow for a decent description of quark mixing by generating the Cabibbo angle at leading order. The symmetry breaking pattern indicated here might provide an interesting opportunity for model building.

\section{Dirac neutrinos and the mixing patterns\label{sec:dirac}}
In this section we assume that neutrinos are Dirac particles and ask the question: What is the smallest finite discrete group $G_f$ that can predict experimentally acceptable PMNS and CKM matrix. The residual symmetry group of neutrino masses is no longer restricted to be isomorphic to the Klein group, but may be an arbitrary abelian group. We scan all the abelian subgroups of every discrete group $G_f$ up to the size of 200. The two smallest finite discrete groups that predict experimentally acceptable entries for the quark and lepton mixing angles are of the order of 150 and 162, with the structure of the relevant remnant groups given in Table \ref{tab:dirac}. An exact definition of the groups in terms of 3-dimensional generators is provided in Appendix \ref{sec:app-def-of-subs}, where we restrict ourselves to listing the smallest subgroups for $\{ G_e, G_\nu \}$ and $\{ G_u,G_d\}$ that predict the given values for the PMNS and CKM mixing parameters.

In our previous scan \cite{Holthausen:2012wt} we had assumed that neutrinos are Majorana particles and found only discrete groups with order of 600 and above that can lead to acceptable leptonic mixing pattern. A priori we have no evidence up till now that neutrinos are Majorana particles and by assuming that neutrinos are Dirac particles, we found two discrete groups that are relatively small in size which can predict experimentally acceptable LO mixing angles for quarks and leptons, which from model building perspective are more economical. The CKM prediction can also be ignored if one only looks for smallest discrete flavor group that can yield the experimentally viable leptonic mixing angles with the assumption of Dirac neutrinos.

To be concrete, we will discuss the group $\Delta(6\times 5^2)$ here in some detail and will relegate the remaining groups to the appendix \ref{sec:app-def-of-subs}.\footnote{Note that the group $\Delta(6\cdot 5^2)$ is also discussed in Ref.~\cite{Lam:2013ng}, however the author only searched for the subgroup $Z_2$ in $\Delta(6\cdot 5^2)$, yielding another type of prediction.} The group $\Delta(6\times 5^2)$ may be viewed as generated by 
\begin{align*}
A=(T U(5,1))^4T^2, \qquad B=(U(5,1) T^2)^2 U(5,1).
\end{align*}
After symmetry breakdown to $G_e=\langle A \rangle \cong Z_3$ and $G_\nu=\langle B\rangle \cong Z_{10}$ the PMNS mixing angles of the first line in Table \ref{tab:dirac} are realized. The CKM predictions follow from breakdown to 
\begin{align*}
G_d=\langle A^2B^3A^2B^2\rangle,\qquad G_u=\langle  ABA^2BA^2B^3A \rangle.
\end{align*}
For a definition of the other groups in Table \ref{tab:dirac}, the reader is referred to Appendix \ref{sec:app-def-of-subs}.

From the definition of A and B in terms of generators of $\Delta(6\cdot 10^2)$ it is clear that $\Delta(6\cdot 5^2)$ is a subgroup of $\Delta(6\cdot 10^2)$. Both groups predict the same PMNS matrix, stemming from different remnant symmetries. The group $\Delta(6\cdot 10^2)$ is the smallest group that predicts LO leptonic mixing patterns in 3-sigma region assuming Majorana neutrinos. If we lift this requirement and allow for Dirac neutrinos, the size of $G_f$ is reduced by a factor of 4. This observation suggests that the leptonic mixing pattern has no correlation with the nature of neutrinos (i.e. whether $Z_2\times Z_2$ is a subgroup of $G_f$ or not) but rather the intrinsic representation of the group generators, i.e. different subgroups can give rise to the same mixing patterns, independent of the nature of neutrinos. The same argument also applies for $(Z_9\times Z_3)\rtimes \mathcal{S}_3$ and $\Sigma(3\cdot 3^3)\rtimes Z_2$ as these 
groups are subgroups of $(Z_{18}\times Z_6)\rtimes \mathcal{S}_3$. All the interesting groups in Table \ref{tab:dirac} predict a trivial Dirac CP phase in the leptonic sector, as in Ref.~\cite{Holthausen:2012wt}.

Combining the argument above and the observation in Sec.~\ref{sec:result}, we can draw a general conclusion that only groups of type $(Z_{n}\times Z_{n'})\rtimes \mathcal{S}_3$  can yield experimentally favored LO PMNS matrix if the flavor symmetry group is broken in such a way that residual symmetries of the leptonic masses are still preserved, independent of whether neutrinos are Dirac or Majorana particles. No other (small) finite discrete groups can yield such an equally successful prediction. In addition the LO CKM mixing pattern can be obtained from group of type $(Z_{n}\times Z_{n'})\rtimes \mathcal{S}_3$ if the size of the group is sufficiently large, as we have pointed out in Sec.~\ref{sec:result}.

\section{Towards Quantifying the Predictive Power Of Discrete Groups \label{sec:anarchy}}   
\begin{figure*}[ht]
\begin{center}
\subfloat[$\mathcal{S}_4$]{\includegraphics[width=0.45\textwidth]{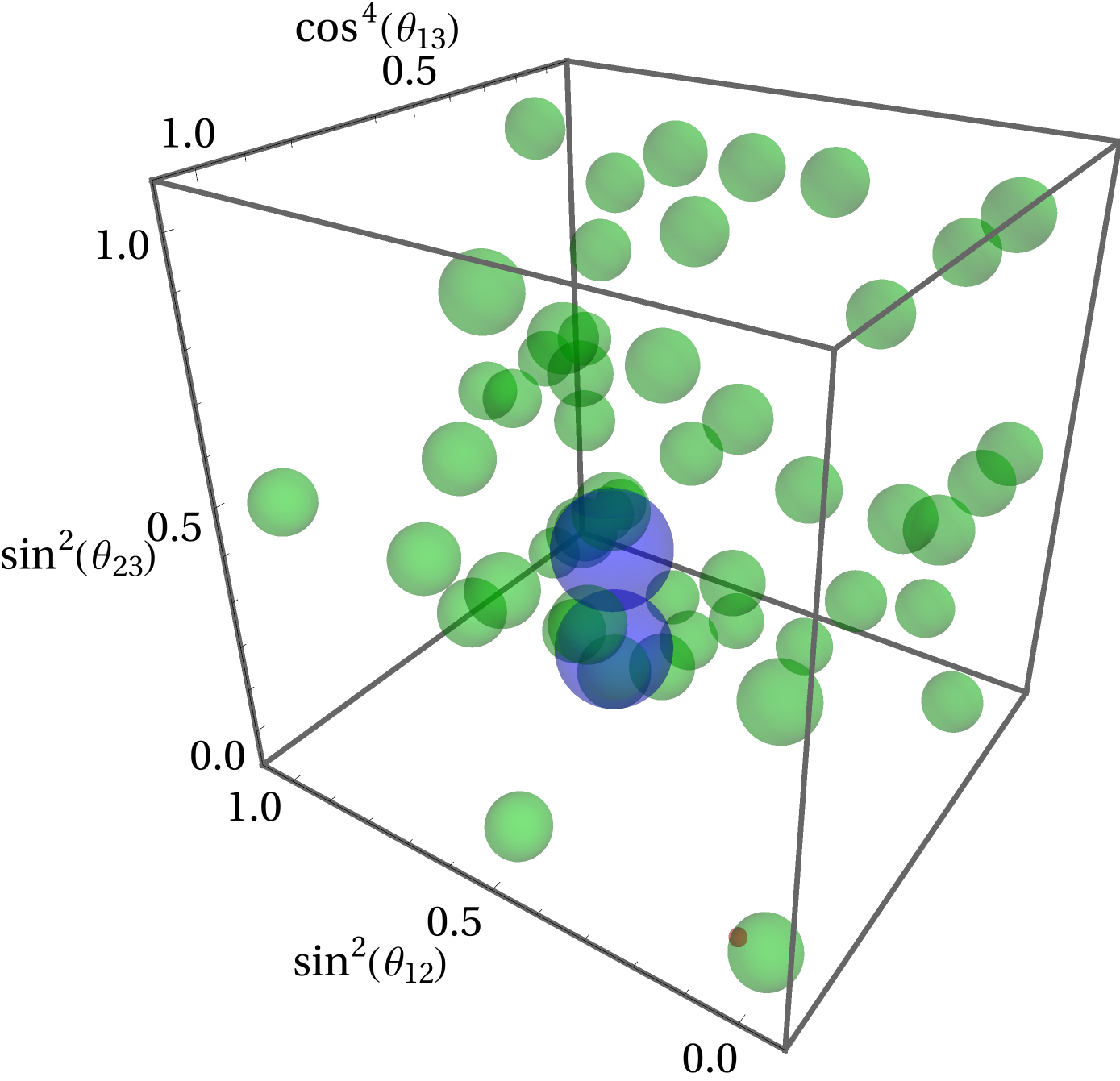}}\hspace{0.05\textwidth}
\subfloat[$(Z_{18}\times Z_6)\rtimes \mathcal{S}_3$]{\includegraphics[width=0.45\textwidth]{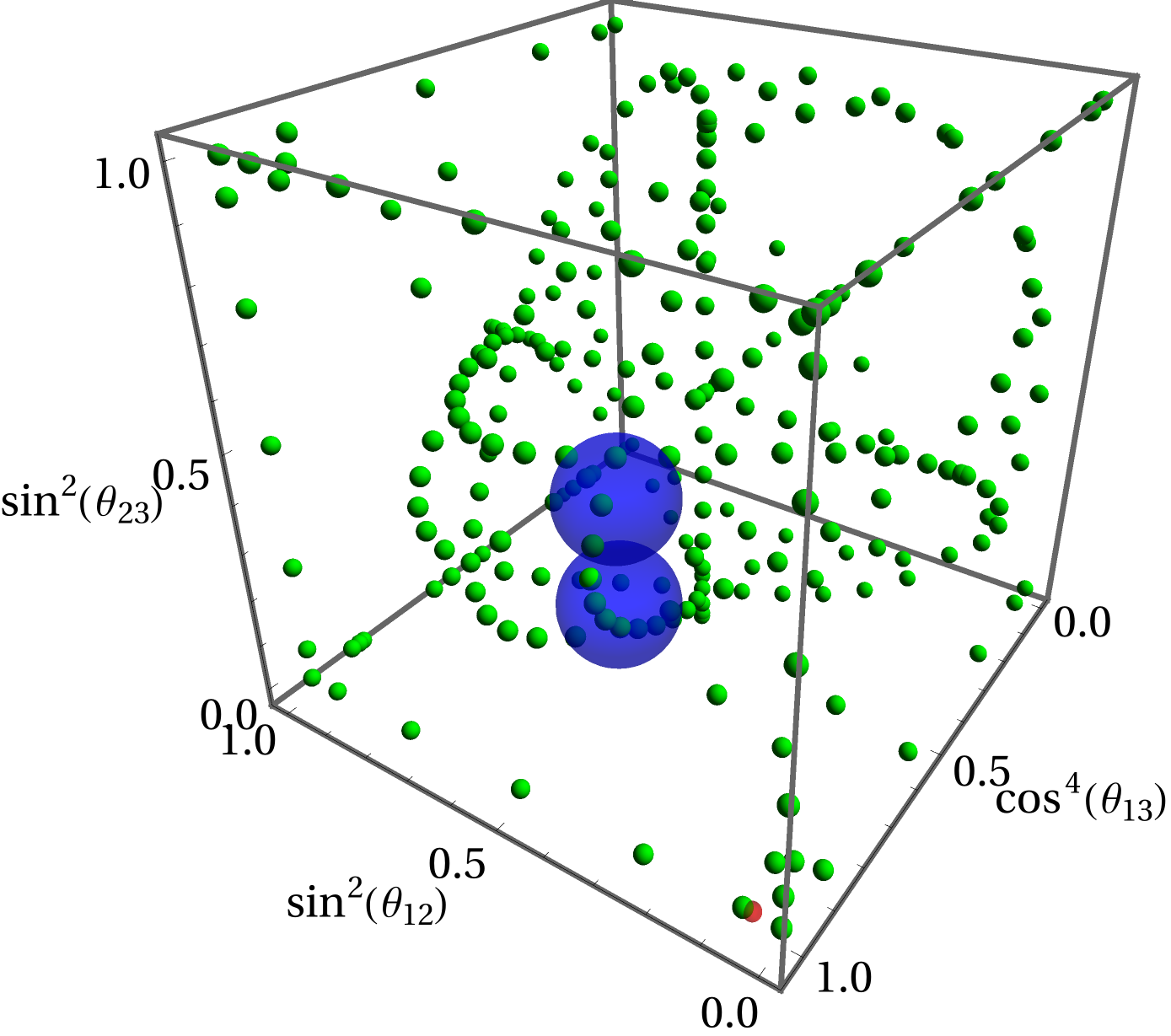}}
\end{center}
\caption{The distribution $\mu(G_f)$ is plotted for groups $\mathcal{S}_4$(left) and $(Z_{18}\times Z_6)\rtimes \mathcal{S}_3$(right). The width of Gaussian Distribution $\sigma$ in 1-sigma deviation is plotted in green. The blue (red) region represents the 3-sigma global fit region for the leptonic (quark) mixing pattern.}
\label{fig:distribution}
\end{figure*}
As mentioned in the introduction, large flavor groups generically have many different abelian subgroups and since in the setup we are considering here the LO mixing pattern is a result of the mismatched remnant symmetries, this implies that for very large flavor groups any mixing pattern should be able to be reproduced. Heuristically, it is therefore clear that one should prefer small flavor groups (which are also less cumbersome from a model builder's view point).  However, we have seen from our scan that only groups that are larger than the order of 100 predict experimentally favored PMNS and CKM mixing pattern at LO. One may wonder what the difference between such a large group and an anarchical \cite{Hall:2000qy,de-Gouvea:2003uq,de-Gouvea:2012fj} drawing of three angular values from the Haar measure.\footnote{See also Ref.~\cite{Hirsch:2001kx,Espinosa:2003yq} for a critical take on anarchy in the lepton sector.}
 
In this section we aim to give a quantitative measure of the predictivity of discrete flavor groups. The scenario we have in mind is the following: we assume the LO quark and/or lepton mixing to be determined from mismatched remnant symmetries, where we take each possible LO mixing pattern to be equally likely. We further assume that NLO corrections are randomly scattered around the LO values. This seems to be well motivated from a model-building perspective as quite often there are a multitude of higher-dimensional operators contributing at NLO order.\footnote{Since in typical models (e.g. \cite{Altarelli:2005uq,Altarelli:2006qy,Holthausen:2011vd,Luhn:2012bc}) these higher dimensional operators do not respect any remnant symmetries this agnostic approach seems warranted. However, it should be stressed that this does not apply for all models and in a particular model the structure of NLO corrections might very well be predictive \cite{King:2011zj}. Such setups usually forbid higher dimensional operators in 
the superpotential; care has to be taken to keep K\"{a}hler corrections under control \cite{Chen:2013aya}.} We discard the comparison of CP phases as the Dirac CP phase in the leptonic sector is not known while the CKM CP phase in general is not predicted in our approach.
  
We will work in the coordinates $c_{13}^4\equiv\cos^4\theta_{13}$, $s_{12}^2\equiv\sin^2\theta_{12}$ and $s_{23}^2\equiv\sin^2\theta_{23}$ for which the invariant Haar measure of $\SU{3}$ is flat. Under the anarchy hypotheses, in this space each point is equally likely $p\mathrm{d}V=\mathrm{d}c_{13}^4\mathrm{d}s_{12}^2\mathrm{d}s_{23}^2$. Without NLO corrections, the discrete group would predict a sum of Delta functions $p\mathrm{d}V=\sum_{i} \delta^{(3)}(\vec{x}-\vec{x}_i)\mathrm{d}c_{13}^4\mathrm{d}s_{12}^2\mathrm{d}s_{23}^2$ centered about the possible LO predictions  $\vec{x}_i=(c_{13}^4, s_{12}^2,s_{23}^2)^T$. Since we expect the NLO corrections to be anarchically distributed around the LO predictions, we smear out the Delta functions to 3-dimensional gaussians $p_f^{(i)}=\exp (\vec{x}-\vec{x}_i)^2/\sigma^2$ centered around the i-th LO mixing with variance given by 
\begin{align}
\sigma^2=\mathrm{Min}(\sigma_{\mathrm{CKM}}^2)+\mathrm{Min}(\sigma_{\mathrm{PMNS}}^2)
\label{eq:variance}
\end{align}
the quadratic sum of the shortest distance between the best fit CKM angles $\vec{x}_\mathrm{CKM}$ and PMNS angles $\vec{x}_\mathrm{PMNS}$ to a LO prediction of the group 
\begin{align}
\mathrm{Min}(\sigma_{\mathrm{CKM/PMNS}})&\equiv \inf_{i} |\vec{x}_i-\vec{x}_\mathrm{CKM/PMNS}|.
\label{eq:min}
\end{align}
The total normalized distribution $p_f$ of a discrete group $G_f$ is given by the sum of all the $p_f^{(i)}$. For illustration in Fig.~\ref{fig:distribution} we show the $p_f$ distribution in the space of $(c^4_{13},s_{12}^2,s_{23}^2)$ for the group $\mathrm{S}_4$ and $(Z_{18}\times Z_6)\rtimes \mathcal{S}_3$. The group $(Z_{18}\times Z_6)\rtimes \mathcal{S}_3$ predicts more mixing patterns than $\mathcal{S}_4$ with smaller covariance $\sigma^2$, as its predicted PMNS matrix values are more accurate at LO.

\begin{figure*}[t]
\centering
\includegraphics[width=.8\textwidth]{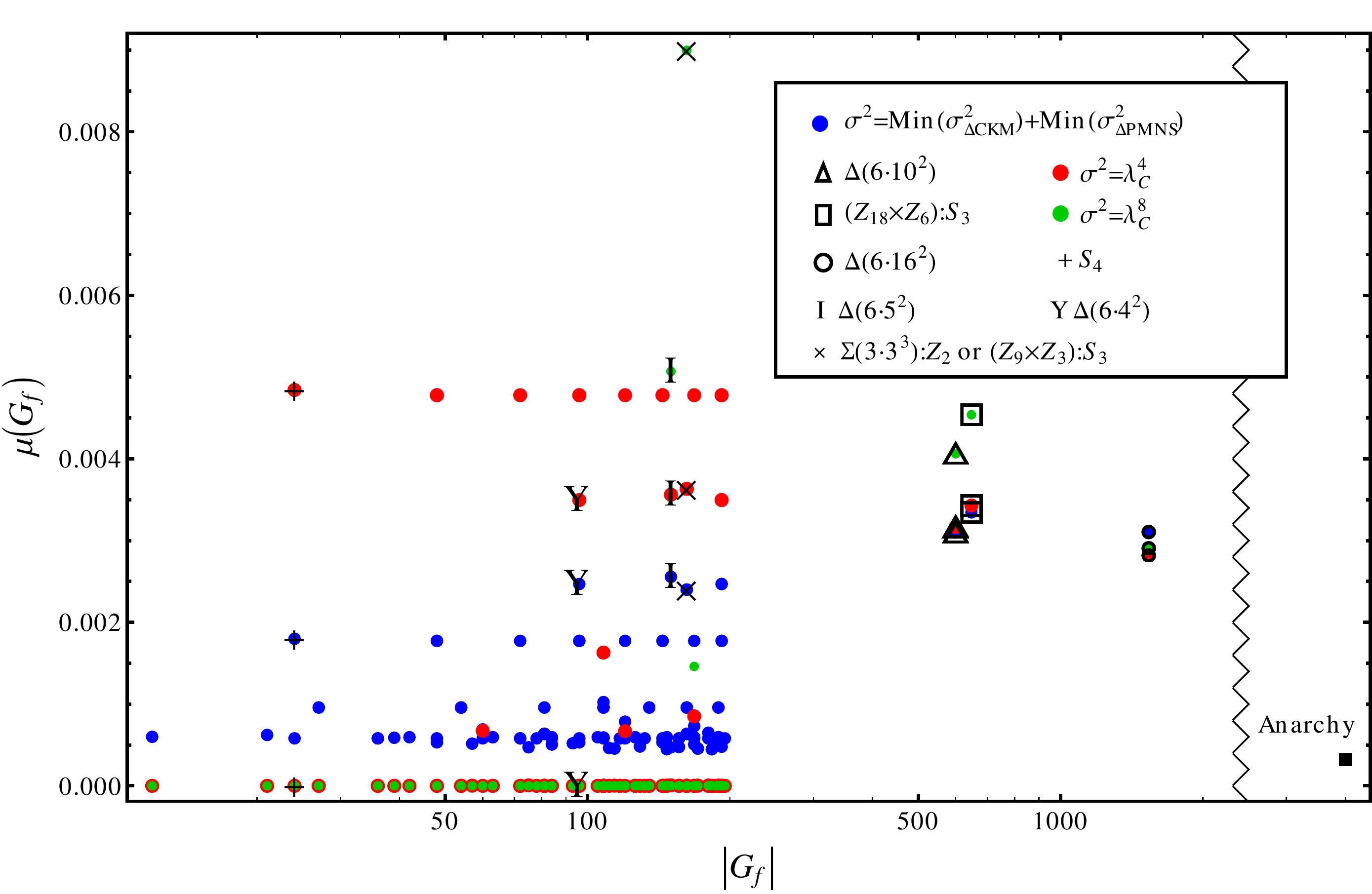} 
\caption[]{The goodness of prediction $\mu(G_f)$ for discrete flavor symmetry groups $G_f$ is plotted. The goodness of prediction for anarchy is represented by a black square in the plot. Groups that are relevant for our analysis are highlighted. See main text for more explanation.}
\label{fig:ckmpmns}
\end{figure*}

As a measure of predictivity we now propose the integration of $p_f$ within the 3-sigma region from global fits 
 \begin{align}
 \mu(G_f) \equiv \int\limits_{V_{\mathrm{exp}}} p_f(c_{13}^4,s_{12}^2,s_{23}^2)\, \mathrm{d}c_{13}^4\mathrm{d}s_{12}^2\mathrm{d}s_{23}^2,
 \label{eq:probab}
 \end{align}
which we interpret as a proxy for the goodness of the mixing angle prediction up to the NLO correction by a particular flavor symmetry group. For example we have $\mu(\mathcal{S}_4)=1.8\times 10^{-3}$ and $\mu((Z_{18}\times Z_6)\rtimes \mathcal{S}_3)=4\times 10^{-3}$. The larger group that needs smaller NLO corrections therefore beats the smaller group with larger NLO corrections -- a result that should not come as a surprise to the reader, who has followed us thus far. 

We can go a step further and apply the measure to anarchy and obtain
\begin{align}
\mu(\mathrm{anarchy})&=\int\limits_{V_{\mathrm{exp}}} \mathbb{1}_{[0,1]^3}\, \mathrm{d}c_{13}^4\mathrm{d}s_{12}^2\mathrm{d}s_{23}^2 \nonumber \\
&=3.22\times 10^{-4},
\label{eq:anarchy}
\end{align}
which might be interpreted as the least predictive theory. Any flavor theory should certainly be more predictive than anarchy.

The result of $\mu(G_f)$ for each discrete group up to the order 200 and some of interesting groups identified by us in Ref.~\cite{Holthausen:2012wt} are plotted with blue points in Fig.~\ref{fig:ckmpmns}\footnote{Some of the higher order groups that yield the same $\mu(G_f)$ as the lower order group contain the same lower order group as their subgroup. For instance the group $\mathcal{S}_4\times Z_2$ and $\mathcal{S}_4 \times Z_3$ yields the same order of $\mu(G_f)$ as the group $\mathcal{S}_4$. }. By this measure the group $(Z_{18}\times Z_6)\rtimes \mathcal{S}_3$ therefore wins the title of the most predictive group smaller than 1536.

\begin{figure*}[ht]
\centering
\includegraphics[width=.8\textwidth]{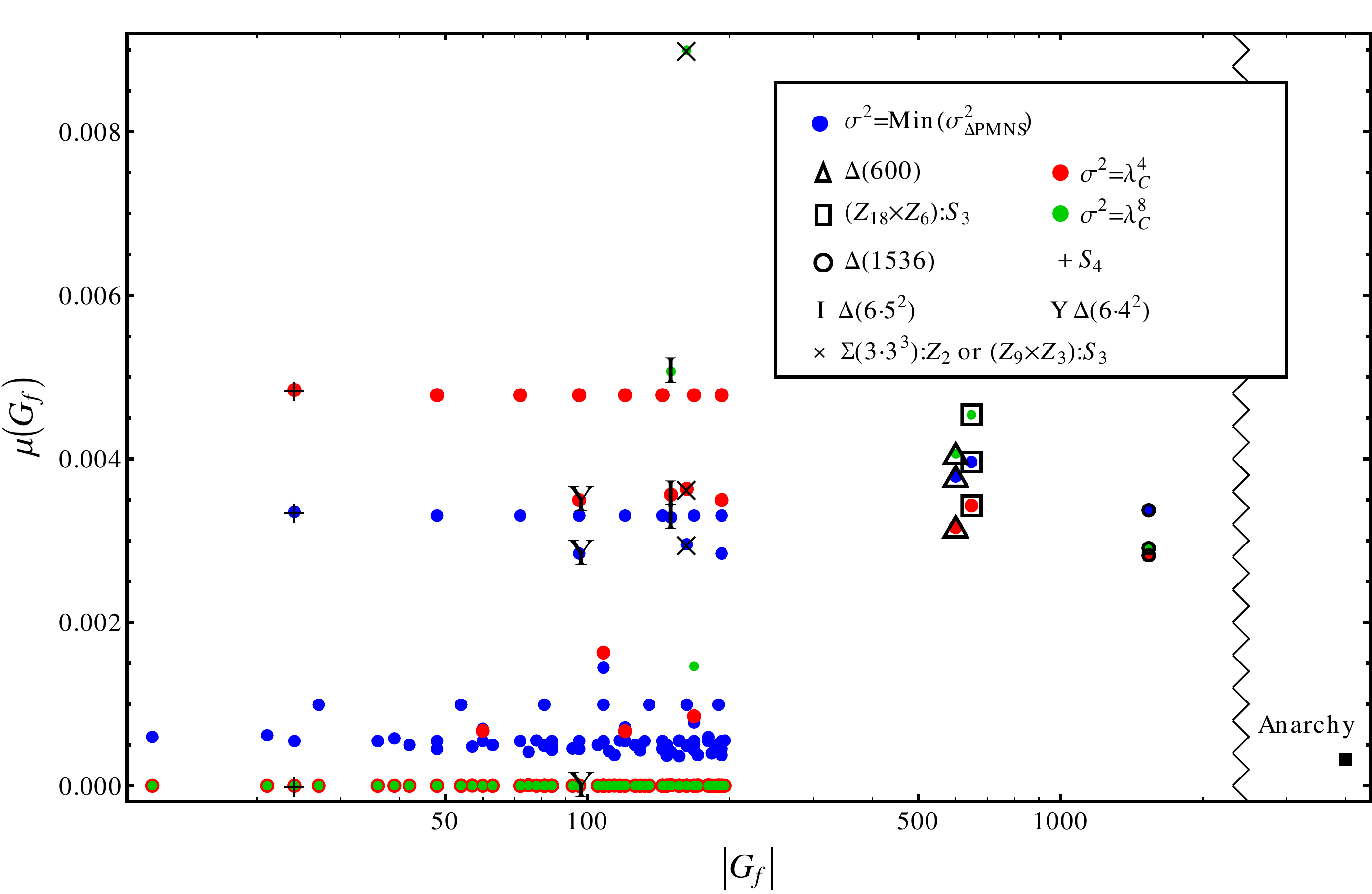} 
\caption[]{The goodness of prediction $\mu(G_f)$ for discrete flavor symmetry groups $G_f$ is plotted, with only the leptonic mixing patterns considered.}
\label{fig:pmnsonly}
\end{figure*} 

Note that the absolute value of $\mu(G_f)$ alone has no intrinsic meaning, rather it is used to compare the goodness of prediction for different flavor groups. A higher value of $\mu(G_f)$ implies a more accurate prediction of mixing angles with smaller size of the group. Groups that do not predict experimentally favored values have smaller values of $\mu(G_f)$. Even though a larger group tends to predict more accurate values of mixing angles, its sizable order would in general reduce the value of $\mu(G_f)$. From Fig.~\ref{fig:ckmpmns} we observe that $\Delta(6\cdot 16^2)$ yields a lower $\mu(G_f)$ value than $(Z_{18}\times Z_6)\rtimes \mathcal{S}_3$ and $\Delta(6\cdot 10^2)$, despite that $\Delta(6\cdot 16^2)$ predicts a more accurate mixing pattern. Ignoring the CKM contributions we can also obtain a similar plot in Fig.~\ref{fig:pmnsonly} by choosing $\sigma^2=\mathrm{Min}(\sigma_\mathrm{PMNS}^2)$. The result of $\mu(G_f)$ for each flavor groups contains the same trend as in Fig.~\ref{fig:ckmpmns}. One 
should note that that by combining the different subgroups of $G_f$ in pairs, we essentially give up the information of the assumption that neutrinos are Majorana, as one needs to pair up only a Klein group with another abelian subgroup if this additional assumption is made.  

The width of the Gaussian distribution defined in Eq.~\eqref{eq:variance} is only one of the possibilities that we can choose. It is believed that the NLO correction of the leptonic mixing angle has to be of the order of Cabibbo angle squared $\sigma=\lambda_C^2$ or the fourth power of Cabibbo angle $\sigma=\lambda_C^4$. We also plotted the result of $\mu(G_f)$ obtained with these assumptions and the only significant change in Fig.~\ref{fig:ckmpmns} and Fig.~\ref{fig:pmnsonly} comes from the group $\mathcal{S}_4$, $\Delta(6\cdot 4^2)$, $\Delta(6\cdot 5^2)$, $(Z_9\times Z_3)\rtimes \mathcal{S}_3$ and $\Sigma(3\cdot 3^3)\rtimes Z_2$. These changes can be understood as the result of higher volume covered by the integration due to more narrow Gaussian width. With $\sigma=\lambda_C^2$, the spread of the Gaussian distribution is larger, hence smaller groups tend to yield higher values of $\mu(G_f)$. On the contrary the Gaussian width is too narrow for $\sigma=\lambda_C^4$, hence only groups that predict very 
accurate LO PMNS matrix will generate a higher $\mu(G_f)$. In fact, $\mu(G_f)$ from anarchy is higher than certain groups, particularly $\mathcal{S}_4$ and $\Delta(6 \cdot 4^2)$. The decreasing value of $\mu(G_f)$ with respect to the increasing size of the group agrees with our naive expectation that higher order groups tend to yield lower value of $\mu(G_f)$ due to more possible combinations of the mixing patterns.

\section{Conclusion \label{sec:conclusion}}
In summary, we have extended our search for discrete symmetry groups that can give an experimentally favored LO prediction for the PMNS and the CKM matrix. With the assumption of Majorana neutrinos, we obtain sizable prediction of LO CKM matrix from groups that predict PMNS matrix in 3-sigma region, as shown in Ref.~\cite{Holthausen:2012wt}. We found a group theoretical reason that explains the emergence of such LO Cabibbo angle, mainly it is due to the structure of $(Z_m \times Z_{m'})\rtimes Z_2$ which is a generic subgroup of $(Z_n \times Z_{n'})\rtimes \mathcal{S}_3$. By relaxing the condition of Majorana neutrinos, we performed a scan of all discrete symmetry groups up to the order of 200 and obtain 3 groups that predict acceptable LO PMNS and CKM matrix.  All 3 groups are subgroups of the groups found in the Majorana case, indicating that mixing pattern predictions are independent of whether neutrinos are Dirac or Majorana particles. We extrapolated our result and concluded that only groups 
that are of the type $(Z_n \times Z_{n'})\rtimes \mathcal{S}_3$ can give experimentally favored values of PMNS (and CKM) matrix, which can provide a new starting point for model building.  

The groups we have found are generally large and prompting us to define a measure to quantify the predictivity of a given flavor group taking into account the size of the group. Our measure $\mu(G_f)$ rewards the smallness of a group while punishing large groups that give many different predictions, depending on the breaking pattern. While this measure is non-unique, it is (to our knowledge) first attempt to quantify more sociological ways of distinguishing fruitful starting points for model building.

\appendix

\section{Definition of Subgroups}
\label{sec:app-def-of-subs}
In Table \ref{tab:generator} we define the generators for groups found in Table \ref{tab:dirac}.
\begin{table*}[tb]
 \centering
 \begin{tabular*}{1\textwidth}{@{\extracolsep{\fill} }ll|l}
 \hline
 $G_f$ & \GAP-Id & Generators of subgroups  \\
 \hline
 $\Delta(6\cdot 5^2)$ & $[150,5]$ & $\langle G_e,G_\nu \rangle=
\left\langle
\left(
\begin{array}{ccc}
 0 & -(-1)^{3/5} & 0 \\
 0 & 0 & -\sqrt[5]{-1} \\
 -\sqrt[5]{-1} & 0 & 0 \\
\end{array}
\right) ,
\left(
\begin{array}{ccc}
 (-1)^{3/5} &0&0 \\
 0&0& \sqrt[5]{-1} \\
 0& \sqrt[5]{-1}&0 \\
\end{array}
\right) \right\rangle
$ \\
&& $\langle G_u,G_d \rangle=
\left\langle
\left(
\begin{array}{ccc}
 (-1)^{3/5} & 0 &0 \\
 0 & 0 & -(-1)^{2/5} \\
 0& -1 & 0 \\
\end{array}
\right) ,
\left(
\begin{array}{ccc}
 (-1)^{3/5} &0&0 \\
 0&0& (-1)^{3/5} \\
 0& -(-1)^{4/5}&0 \\
\end{array}
\right) \right\rangle
$ \\
 \hline
 $\Sigma(3\cdot 3^3)\rtimes Z_2$ & $[162,10]$ & $\langle G_e,G_\nu \rangle=
\left\langle
\left(
\begin{array}{ccc}
 0 & 0& -1 \\
 0 & -\sqrt[3]{-1} &0\\
 -1 & 0 & 0 \\
\end{array}
\right) ,
\left(
\begin{array}{ccc}
 0&0& (-1)^{2/3} \\
 (-1)^{2/3} &0 &0 \\
 0& -\sqrt[3]{-1}&0 \\
\end{array}
\right) \right\rangle
$ \\
&& $\langle G_u,G_d \rangle=
\left\langle
\left(
\begin{array}{ccc}
 0 & 0& -1 \\
 0 & -\sqrt[3]{-1} &0\\
 -1 & 0 & 0 \\
\end{array}
\right) ,
\left(
\begin{array}{ccc}
 0 & 0& \sqrt[3]{-1} \\
 0 & -1 &0\\
 -1 & 0 & 0 \\
\end{array}
\right) \right\rangle
$ \\  \hline
 $(Z_9\times Z_3)\rtimes \mathcal{S}_3$ & $[162,12]$ & $\langle G_e,G_\nu \rangle=
\left\langle
\left(
\begin{array}{ccc}
 0 & 0& (-1)^{5/9} \\
 0 & (-1)^{8/9}-(-1)^{5/9} &0\\
 (-1)^{5/9} & 0 & 0 \\
\end{array}
\right) \color{white}\right \rangle \color{black},$ \\
&&\hspace{1.3cm} $
\color{white}\left\langle \color{black}\left(
\begin{array}{ccc}
 0&0& (-1)^{5/9}-(-1)^{8/9} \\
 -(-1)^{5/9} &0 &0 \\
 0& -(-1)^{5/9}&0 \\
\end{array}
\right) \right\rangle
$ \\
&& $\langle G_u,G_d \rangle=
\left\langle
\left(
\begin{array}{ccc}
 0 & 0& (-1)^{5/9} \\
 0 & (-1)^{8/9}-(-1)^{5/9} &0\\
 (-1)^{5/9} & 0 & 0 \\
\end{array}
\right) \color{white}\right \rangle \color{black}, $ \\
&&\hspace{1.3cm} $ \color{white}\left \langle \color{black} 
\left(
\begin{array}{ccc}
 0 & 0& (-1)^{8/9}-(-1)^{5/9} \\
 0 & (-1)^{8/9}-(-1)^{5/9} &0\\
 -(-1)^{8/9} & 0 & 0 \\
\end{array}
\right) \right\rangle
$ \\  \hline
$(Z_9\times Z_3)\rtimes \mathcal{S}_3$ & $[162,14]$ & $\langle G_e,G_\nu \rangle=
\left\langle
\left(
\begin{array}{ccc}
 0 & 0& (-1)^{5/9} \\
 0 & -(-1)^{8/9} &0\\
 (-1)^{5/9} & 0 & 0 \\
\end{array}
\right) \color{white}\right \rangle \color{black},$ \\
&&\hspace{1.3cm} $
\color{white}\left\langle \color{black}\left(
\begin{array}{ccc}
 0& \sqrt[9]{-1}-(-1)^{4/9} & 0 \\
 0& 0& (-1)^{4/9} \\
 \sqrt[9]{-1}-(-1)^{4/9}&0 &0\\
\end{array}
\right) \right\rangle
$ \\
&& $\langle G_u,G_d \rangle=
\left\langle
\left(
\begin{array}{ccc}
 0 & 0& (-1)^{5/9} \\
 0 & -(-1)^{8/9} &0\\
 (-1)^{5/9} & 0 & 0 \\
\end{array}
\right) \color{white}\right \rangle \color{black}, $ \\
&&\hspace{1.3cm} $ \color{white}\left \langle \color{black} 
\left(
\begin{array}{ccc}
 0 & 0& -(-1)^{8/9} \\
 0 & -(-1)^{8/9} &0\\
 (-1)^{8/9}-(-1)^{5/9} & 0 & 0 \\
\end{array}
\right) \right\rangle
$ \\  \hline
 \end{tabular*}
\caption{Generators for $\{G_e,G_\nu\}$ and $\{G_d,G_u\}$ that predicts the experimentally favored mixing angles in Table \ref{tab:dirac}.}
\label{tab:generator}
\end{table*}

\vspace*{2mm}

\noindent {\bf Acknowledgements:} We would like to thank Michael A. Schmidt, Claudia Hagedorn and Werner Rodejohann for useful comments on the manuscript. We also want to thank Yusuke Shimizu for useful discussions. K.S.L. acknowledges support by the International Max-Planck Research School for Precision Tests of Fundamental Symmetries. 

\bibliographystyle{apsrev4-1}
\bibliography{biblio2}
\end{document}